# A BI-SCHEDULER ALGORITHM FOR FRAME AGGREGATION IN IEEE 802.11N

Vanaja Ramaswamy[1], Abinaya Sivarasu[2], Bharghavi Sridharan[3] and Hamsalekha Venkatesh[4]

[1, 2, 3, 4] Department Of Computer Science and Engineering,
Sri Venkateswara College Of Engineering,
Sriperumpudur taluk, Tamilnadu, India

## ABSTRACT

*IEEE 802.11n mainly aims to provide high throughput, reliability and good security over its other previous standards. The performance of 802.11n is very effective on the saturated traffic through the use of frame aggregation. But this frame aggregation will not effectively function in all scenarios. The main objective of this paper is to improve the throughput of the wireless LAN through effective frame aggregation using scheduler mechanism. The Bi-Scheduler algorithm proposed in this article aims to segregate frames based on their access categories. The outer scheduler separates delay sensitive applications from the incoming burst of multi-part data and also decides whether to apply Aggregated - MAC Service Data Unit (A-MSDU) aggregation technique or to send the data without any aggregation. The inner scheduler schedules the remaining (delay-insensitive, background and best-effort) packets using Aggregated-MAC Protocol Data unit (A-MPDU) aggregation technique.*

## 1. INTRODUCTION

IEEE 802.11n is the most popularly used wireless *Local Area Network (LAN)* technology today. It is one of the superior standards when compared to its previous standards like IEEE 802.11 a/b/g through its ability to provide maximum of 600 Mbps [4] data rate. This is due to the enhancements provided in IEEE 802.11n like Multiple Input Multiple Output (MIMO) and channel bonding [1] in PHY (Physical) layer and other enhancements in Medium Access Control (MAC) layer. The MIMO technology in PHY layer uses multiple antenna elements to send multiple copies of the data there by significantly increasing the data rate. It uses 40 MHz channel for transmission. The MAC layer enhancements are frame aggregation, block acknowledgement, reduced inter frame spacing and allowing transmission in reverse direction. Recently, many 802.11n based products are popularly used in many user devices. But the performance of these products is not up to the expectation due to various constraints in the real time. The fairness of the IEEE 802.11n is not very effective in case of unsaturated traffic.

This paper proposes a scheduler for the MAC layer which divides the packet based on Access Category (AC) to provide different QoS services for different category of data. It also decides between whether to use frame aggregation or not based on the urgency of the data. This provides fairness to the traffic for off time traffic.

DOI : 10.5121/ijwmn.2013.5607 91

## 2. RELATED WORK AND LITERATURE SURVEY
### 2.1 PHY and MAC layer enhancements

The important enhancements in IEEE 802.11n which improved the performance of the WLAN are briefed here.

### 2.1.1 Multiple Input Multiple Output (MIMO)

IEEE 802.11n uses MIMO technology to improve the data rate up to 600 Mbps [10]. MIMO uses multiple antennas for transmission and reception for transmission of the same data at phases. It defines different M $X$ N configurations from 1 $X$ 1 to 4 $X$ 4 [5]. As the number of antennas for transmission and reception increases, the performance of the channel increases significantly. MIMO operates in two different modes: SDM and STBC.

- In Spatial Division Multiplexing (SDM), the output stream is divided into multiple sub streams which are then spatially multiplexed and transmitted from multiple antennas. At the receiving side, these streams are demultiplexed and recovered.
- In Space Time Block Coding (STBC), the same output stream is transmitted over multiple antennas simultaneously. At the receiving side, the multiple streams with different strength are compared to recover the original stream.

| 802.11 | Freq (Hz) | Through put (Mb/s) | Max Bit rate (Mb/s) | Modulation | $r_{in}$ (m) |
|---|---|---|---|---|---|
| A | 5 | 23 | 54 | OFDM | ~35 |
| B | 2.4 | 4.3 | 11 | DSSS | ~38 |
| G | 2.4 | 19 | 54 | OFDM | ~38 |
| N | 2.4, 5 | 74 | 248 | OFDM | ~70 |

Table 2.1- 802.11n standard specification [2]

### 2.1.2 Channel bonding

It uses a 40MHz channel which means greater bandwidth which ultimately leads to better performance in WLAN. The 40MHz channel [5] behaves like a combination of two 20 MHz channel. It can also operate in two modes: 20MHz mode and 40MHz mode. The 20MHz [5] mode is used if the bandwidth availability is limited. The 40MHz channel provides two times more desirability than 20MHz channel.

### 2.2 MAC layer enhancement

Frame aggregation is a feature first introduced in 802.11e Wireless LAN where two or more frame is sent in a single transmission to increase the throughput. Frame aggregation reduces overhead since it transmits multiple frames as a single aggregate frame it reduces traffic

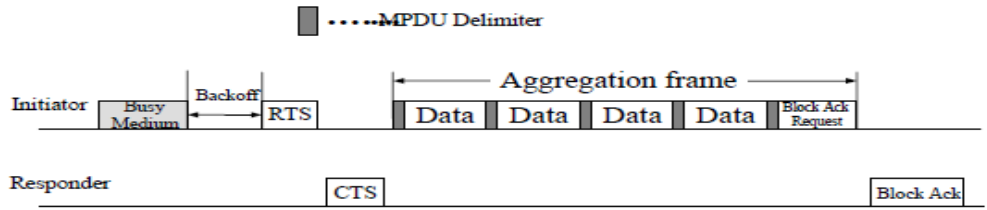

Fig 2.1- Frame Aggregation method [9]

There are three types of frame aggregation techniques- MSDU, MPDU and Two Level Frame Aggregation.

### 2.2.1 MAC Service Data Unit (MSDU)

This is the unit of transmission used at the MAC layer for data which is received from the upper layer. The data received from the logical link control (LLC). This LLC sub layer is present above the medium access control (MAC) sub layer in a protocol stack.

### 2.2.2 Aggregated MAC Service Data Unit (A-MSDU)

Multiple MAC service data unit are combined to form the payload of a MPDU packet. The entire aggregate packet will have a single frame header and these packets are destined to the same client and should have same access category. The main motivation of A-MSDU technique is the smaller size of Ethernet header when compared to 802.11 header. Hence multiple Ethernet frames can be combined to form A-MSDU. Since individual sub frames don't have frame checksum, selective retransmission of corrupted sub frames will not be possible. Sub frames will have same traffic identifier (TID) and sequence number [2]. For constructing an A-MSDU there is some limitations to follow such as, every MSDUs should have the same traffic identifier value. The sustained time of A-MSDU should be equal to the maximum sustained time of its constituent elements. In the sub frame header the values of destination address and sender address should match with the values of receiver address and the transmitter address in the MAC header.

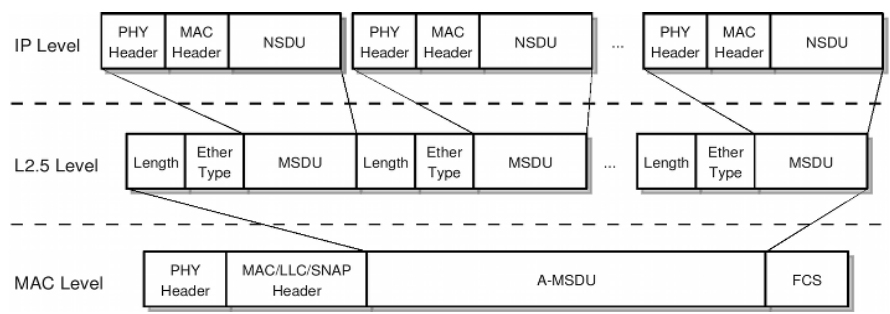

Fig 2.2- Format of A-MSDU [10]

The maximum frame length allowed in A-MSDU scheme is 3kb or 7kb [2]. Every sub frame consists of a header these headers is followed by packets that arrived from logical link control and 0 – 3 bytes of padding [10]. The padding size depends on each sub frame but the last one must be a multiple of four bytes [9].

### 2.2.3 MAC Protocol Data Unit (MPDU)

MAC protocol data units are the frames passed from the MAC layers into the PHY layer. This MPDU mechanism is more efficient with high data rates because of block acknowledgement. This block acknowledgement allows each of the data aggregated frames to be acknowledged individually and if error occurred it is retransmitted.

### 2.2.4 Aggregated MAC Protocol Data Unit (A-MPDU)

Multiple MAC protocol data unit with common PHY header are aggregated as a single A-MPDU. The size of sub frames should be multiple of 4 bytes otherwise length is adjusted by adding padding bytes [9]. Each PDU in the frame has its known header and cyclic redundancy check so in case of transmission error it is sufficient to send only the lost PDU thereby improving the throughput [3]. A-MPDU functions after the encapsulation process of MAC header here the matching traffic identifier is not considered but all the MPDUs present within an A-MPDU should be addressed to the same receiver address. Since there is no holding time to form A-MPDU the number of MPDUs to be aggregated depends on the packets that are already present in the transmission queue.

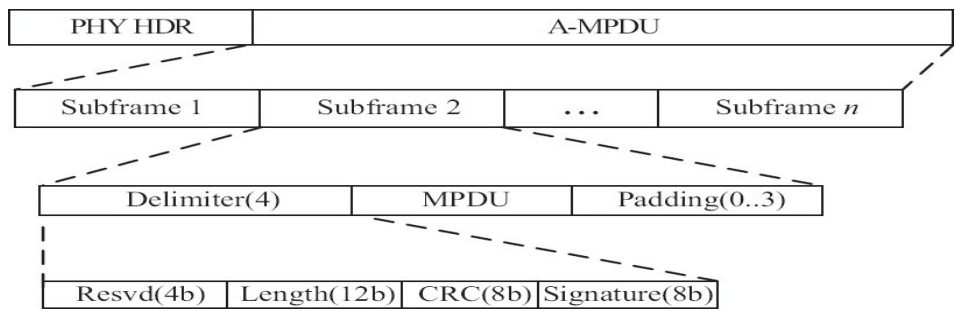

Fig 2.3- Format of A-MPDU [10]

Before MPDU a set of delimiters are added and in its tail padding bits varied from 0 – 3 bytes are added to define the length and position inside the aggregated frame the delimiter header is used. For determining the structure of A-MPDU the usage of MPDU delimiters and PAD bytes are must. After the A-MPDU is received it checks for the MPDU delimiter for any errors in cyclic redundancy check. If the value is correct then the MPDU is extracted else it continues until it reaches the end of physical layer service data unit

### 2.3 Block acknowledgement

The block acknowledgement is a single acknowledgement frame that acknowledges multiple received frames.802.11n allows a receiver to cumulatively acknowledge multiple frames with a single Block Acknowledgement [9].

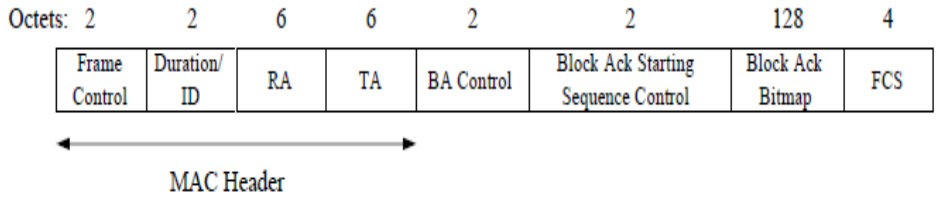

Fig 2.4 - Block acknowledgement format

To confirm the frame delivery block acknowledgment minimizes the number of ACKs that a receiver must send to a transmitter.

| Access Category | Frame Aggregation | Reasons |
|---|---|---|
| Voice | NA | Voice packets are delay-sensitive. These packets need strict latency, jitter and bandwidth requirement [8] use of aggregation degrade the performance of voice packets since these has to wait for other packets. |
| Video | A-MPDU | Video packets have strict bandwidth demand but loser latency in jitter demand [8]. Video packets have high data rate so waiting for other packets exceed the maximum tolerable delay [7]. |
| Best effort and Background | A-MPDU | Delay insensitive and no bandwidth requirements [8] making the packets of this access category will not have any impact on their performance. |

Table 2.2 -Selecting required Aggregation Scheme

## 3. MOTIVATION FOR THE PROPOSED WORK (Issues in frame aggregation)

Generic frame aggregation cannot be applied for all access categories because of different QoS requirements [8]. Frame aggregation is unable to function effectively in unsaturated traffic. As one side, the sender has to wait for more frames which deliberately increases the latency there by affecting delay sensitive packets. Other choice is, it is forced to send the available packet which does not use the bandwidth efficiently. In the above cases, the advantage gained by the frame aggregation is lost.

## 4. PROPOSED WORK

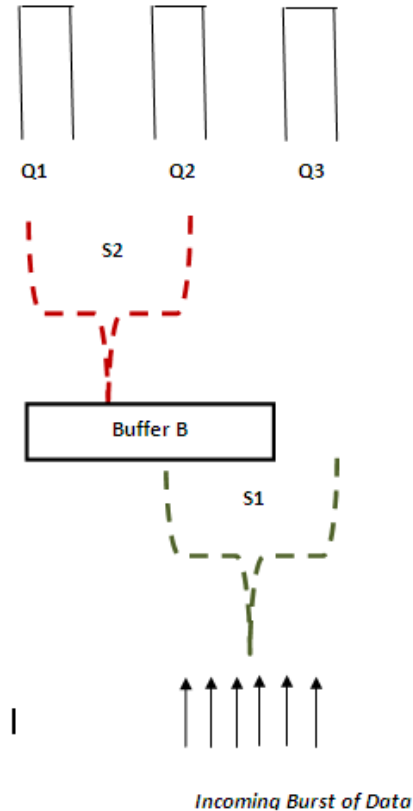

1. **Q1**-Stores Audio packets. i.e. (Delay sensitive packets). We cannot apply frame aggregation to those packets. MSDU Concept has to be used in this case.
2. **Q2**- Stores Video packets. We can apply frame aggregation to those packets.
3. **Q3**- Background/ Best effort packets (delay insensitive packets), which can be sent anytime but will not affect the performance.
4. **S1**- Scheduler 1 (**Outer Scheduler**)
5. **S2**- Scheduler 2 (**Inner Scheduler**)

Fig 4.1- Bi-Scheduler Implementation Using Queues

### 4.1 Working of Bi-Scheduler Technique for Throughput improvement in Frame Aggregation

The idea is based on the fact that since Audio packets are delay sensitive and have low data rate, so making them wait in the queue will introduce lot of delay in network and degrades its performance. Thus, A-MSDU technique must be applied to such packets in order to increase their throughput. Hence frame aggregation (A-MSDU) can be effectively applied only to improve the throughput of high data rate packets. Therefore, making such types of packets wait in the queue (to attain the desired size for aggregation) does not significantly affect the throughput of the network. While Background and Best Effort packets are delay insensitive and making them wait will not have any impact in the through put of the network.

1. Initially, the **Outer Scheduler (S1)** separates the *delay sensitive* frames i.e. Voice/ Audio frames from the incoming burst of multipart data based on their Access Category (AC). These frames are directly enqued into Queue Q1, where they wait for the other frames of same AC to arrive in order to form A-MSDU. A timer is also maintained separately for this queue in order to minimize the delay that is deliberately introduced in this process. If the timer for this queue expires, automatically the contents of Q1 will be sent to respective destination irrespective of the number of frames available at that time.

2. The remaining frames i.e. Video and Background/ Best-Efforts are stored temporarily in Buffer B1 of **Inner Scheduler (S2)**. S2 then separates Video frames from Best-Efforts frames and enques them in queue Q2, where they are aggregated to form A-MPDU packets. The rest of the frames are enqued in Q3. A common timer is also maintained for Q2 and Q3.

   a. When Q2 has reached the aggregated size before the timer expires, it's transferred to the respective destination.

   b. If Q2 has not reached the aggregated size before the timer expires, then that many remaining packets from Q3 are taken to attain the desired size and sent to the destination.

3. If, simultaneously both Q1 and Q2 have reached the desired size before the timer expires, then our algorithm gives preference to Audio packets i.e. contents of Q1 will be sent first, since they are highly delay sensitive packets. After, resetting Q1's timer, Q2 contents will be sent to the respective destination.

Fig 4.2 Bi-Scheduler Algorithm's Flowchart

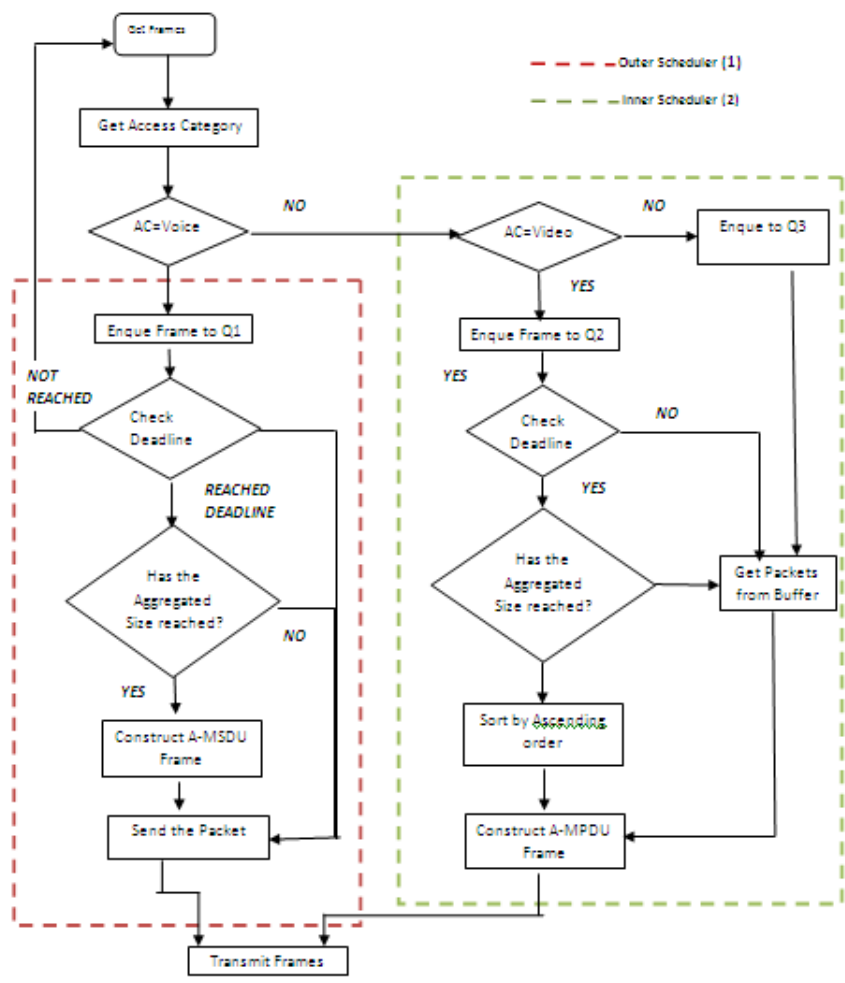

International Journal of Wireless & Mobile Networks (IJWMN) Vol. 5, No. 6, December 2013

## 5. CONCLUSION

The research work proposes a new Bi-Scheduler Mechanism to improve the throughput of IEEE 802.11n WLAN using efficient frame aggregation technique. Depending on the Access Category (AC), the proposed scheduler decides whether to implement A-MSDU or A-MPDU mechanism, in order to efficiently transmit the frames to the destination. The future work will be to simulate and experimentally verify the proposed scheduler mechanism effectively.

## ACKNOWLEDGEMENTS

We would like to acknowledge and thank the Defence Research Development Organisation (DRDO), New Delhi, India for granting us extra mural research funds for carrying out the research work is a part of the research project titled "Network Throughput Enhancement in distributed using Cross layer optimisation".

We would also like to thank the Head of the Department of Department of CSE and other faculty members who gave us necessary support to implement this project.

**Authors**

**Ms Vanaja Ramaswamy**

Assistant professor, Department of Computer Science and Engineering, Sri Venkateswa ra college of Engineering, Sriperumpudur Taluk, Tamilnadu,IndiaCo-Principal Investigator of DRDO Extra mural Research Project.Area of Interest : Operating System , Data structures and Computer Networks

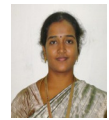

**Hamsalekha Venkatesh**

Final Year Student (B.E.), Department of Computer Science and Engineering Sri Venkateswara College Of Engineering. Sriperumpudur Taluk, Tamilnadu, India,Area of Interest: Computer Networks an d Graph Theory.

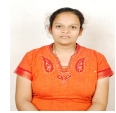

**Bhargavi Sridharan**

Final Year Student (B.E.), Department of Computer Science and Engineering S ri Venkateswara College Of Engineering. Sriperumpudur Taluk, Tamilnadu, India,Area of Interest:  Computer Networks and object oriented programming.

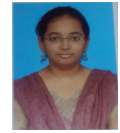

**Abinaya sivarasu**

Final Year Student (B.E.), Department of Computer Science and Engineering Sri Venkateswara College Of Engineering. Sriperumpudur Taluk, Tamilnadu, India,Area of Interest: Computer Networks and operating system

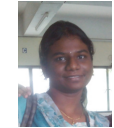